\documentclass[a4paper]{aa}
\usepackage{astron,graphics,psfig,times}
\def\tento#1{$\times$10$^{#1}$}
\def\mtento#1{\mathrel{\times10^{#1}}}
\def\SNII{{\rm SN}$\;\!\!${\sc ii}\,\,}

\begin{document}

   \thesaurus{03     % A&A Section 3: Extragalactic astronomy
              (11.01.2;  % Galaxies: active
	       11.09.1 NGC~4410;  % Galaxies: individual
               11.09.2;  % Galaxies: interactions
	       11.19.3;  % Galaxies: starburst
               13.25.2)} % X-rays: galaxies

   \title{ROSAT X-ray observations of the interacting pair of galaxies\\
	  NGC~4410: evidence for a central starburst}

%   \subtitle{}

   \author{D. Tsch\"oke
          \inst{1}
          \and
          G. Hensler
          \inst{1}
	  \and
	  N. Junkes
	  \inst{2}
          }

   \offprints{D. Tsch\"oke}

   \institute{Institut f\"ur Theoretische Physik und Astrophysik, 
              Universit\"at Kiel, D-24098 Kiel, Germany\\
              email: tschoeke@astrophysik.uni-kiel.de
         \and
             Max-Planck-Institut f\"ur Radioastronomie, Auf dem H\"ugel 69,
             D-53121 Bonn, Germany\\
             }

   \date{Received 1998 July 20 / Accepted 1998 November 6}

   \titlerunning{ROSAT X-ray observations of NGC~4410}
   \maketitle

   \begin{abstract}

We present X-ray observations of the interacting pair of galaxies NGC~4410a/b with the ROSAT HRI and PSPC. The ROSAT HRI images reveal a point-like source corresponding to NGC~4410a and an X-ray halo, extending 10\arcsec\ from the nucleus toward the southeast. The halo emission accounts for $\sim$1/3 ($L_{\mathrm X}$ = 1.3\tento{41} erg s$^{-1}$ in the 0.1 - 2.4~keV ROSAT band) of the total X-ray emission detected from NGC~4410a. The spectrum of the total X-ray emission from NGC~4410a can be fitted at best with a two-component emission model, combining a Raymond-Smith spectrum with $T$ = 10$^7$ K and a power-law spectrum ($\Gamma$ = 2.2). The fraction of the thermal component to the total flux within this model is 35\%, supporting the results of the HRI observation. The total unresolved X-ray luminosity detected with the PSPC amounts to 4\tento{41} erg s$^{-1}$ in the ROSAT PSPC band (for $D$ = 97 Mpc). 

A preferable explanation might be the interaction between the two galaxies causing a circumnuclear starburst around the active nucleus in the peculiar late type Sab galaxy NGC~4410a. As a cumulative effect of exploding stars a superbubble forms a bipolar outflow from the galactic disk with an expansion time of $\sim$8 Myr. The central source injects mechanical energy at a constant rate of a few times 10$^{42}$ erg s$^{-1}$ into the superbubble. 

A HST WFPC2 image decovers that NGC~4410a is seen almost face-on so that only the approaching outflow is visible. The HRI contours reveal an elongation of this outflow indicative for a faint X-ray ridge below the detection limit ($L_{\mathrm X} \le$ 1.4\tento{39} erg s$^{-1}$) toward the neighbouring galaxy NGC~4410b caused by its tidal interaction. NGC~4410b houses only a faint X-ray source ($L_{\mathrm X} \le$ 3.8\tento{39} erg s$^{-1}$).

   \keywords{Galaxies: interactions -- Galaxies: active -- 
	     Galaxies: starburst -- Galaxies: individual: NGC~4410 -- 
	     X-rays: galaxies}

   \end{abstract}

%________________________________________________________________

\section{Introduction}

The pair of interacting galaxies NGC~4410a/b belongs to a group of 11 members (Hummel et al. 1986, hereafter HKG86) which are located behind the Virgo cluster. It consists of a peculiar Sab (NGC~4410a) (Thuan \& Sauvage 1992) and an E galaxy (NGC~4410b) (HKG86) located in east-west direction and separated by 18\farcs7 (8.8 kpc at a distance of 97 Mpc) (Mazzarella \& Boroson 1993, hereafter MB93). NGC~4410a is located at RA = 12$^\mathrm{h}$ 26$^\mathrm{m}$ 27\fs9, Dec = +09\degr\ 01\arcmin\ 18\arcsec\ (J2000). Spectral analysis of the nuclei of both components led to the classification as two LINERs (Mazzarella et al. 1991; Bicay et al. 1995; Thuan \& Sauvage 1992). In the eastern component NGC~4410b a supernova type {\sc i} has been detected in 1965 (SN 1965 A) from which Turatto et al. (1989) derived a distance of 139 Mpc. The system has a radial velocity of about 7300 km s$^{-1}$ (Mazzarella et al. 1991; Batuski et al. 1992; Thuan \& Sauvage 1992; MB93). With a Hubble constant of 75 km s$^{-1}$ Mpc$^{-1}$ this leads to a distance of 97 Mpc. In this paper we apply the latter value to all distance-dependent parameters, like e.g. luminosities etc. A distance of 97 Mpc for NGC~4410 results in an absolute length scale of 470 pc~arcsec$^{-1}$. 

HKG86 found a spatial coincidence of the western optical component NGC~4410a with a radio point source of luminosity  $L_{\mathrm R} \approx$10$^{39}$ erg s$^{-1}$ embedded in a strong, extended radio source around NGC~4410a with a total radio luminosity $L_{\mathrm R} \approx$7\tento{40} erg s$^{-1}$.

The phenomenon of interaction between galaxies is closely related to the occurrence of starbursts (SBs): A large fraction of interacting galaxies ($\sim$70\%) exhibits typical characteristica of SBs (Bushouse 1986) and vice versa. It is also striking that the majority of infrared-bright galaxies shows evidence for recent interactions as indicated by the presence of close neighbours, or by their disturbed morphology and tidal tails (Joseph et al. 1984; Lonsdale et al. 1984; Telesco 1988). The fraction of mergers increases drastically in the infrared luminosity range $L_{IR}$ from 10$^{10} L_{\sun}$(12\%) to 10$^{12} L_{\sun}$(95\%) (Sanders \& Mirabel 1996). Although a large IR luminosity is not necessarily a tracer for enhanced star formation, it is one of the typical features for SB galaxies.

Norman \& Scoville (1988) demonstrated that gas inflow from the galactic disk toward the central region, as required for building a central SB, will be triggered by a non-axialsymmetric gravitational potential like in interacting galaxies. They argued that perturbed orbits of gas clouds caused by an encounter lead to enhanced cloud collisions within the galactic central region and trigger the formation of massive stars. An alternative model for triggering a central SB was given by Jog \& Das (1992) and Jog \& Solomon (1992). The infall of giant molecular clouds (GMCs) into an intercloud medium with a higher mean pressure in the central region drives a radiative shock into the GMCs and ignites the SB.

The inflow of gas from the galactic disk is not only required for a central SB but also for fuelling an AGN. It is still under debate whether SBs are progenitors of AGN (Norman \& Scoville 1988) or whether both are different physical processes. Weedman (1983) proposed that if a large number of massive stars form fast in a small central volume, the compact stellar remnants from these could act as accretors. Also the stellar dynamical merging of a dense cluster of massive stellar remnants would plausibly form a blackhole nucleus (Rees 1984).

The nuclei of several nearby galaxies, like NGC~1068, NGC 1097, and NGC~7469, can be resolved into a central AGN and a circumnuclear SB ring (Keel 1985; P\'erez-Olea \& Colina 1996, hereafter PC96). Some host galaxies of quasars also exhibit evidence for SBs, such as 3C\,48 (Stockton 1990).

Assuming an evolution from starbursts to AGNs during the interaction of galaxies one would expect more AGNs with the age of the merger process. Recent observations of ultraluminous IR galaxies with the infrared satellite ISO however do not show any obvious tendency of an increasing fraction of AGNs within interacting galaxies with advanced merging process (Lutz et al. 1998; Genzel et al. 1998). Half of the observed galaxies reveal both an AGN and starburst activity. It seems more likely that more local and shorter term conditions like time-dependent compression of the circumnuclear interstellar gas, the accretion rate onto the central black hole, and the associate radiation efficiency determine AGN or starburst dominated luminosities.

In this paper we present the spectral and imaging X-ray properties of NGC~4410 observed by the ROSAT Position Sensitive Proportional Counter (PSPC) and High Resolution Imager (HRI), respectively. We are able to resolve the two optical components in the ROSAT HRI image. The paper is structured as follows: In the next section we are considering the X-ray observations in general, detecting and trying to identify the emergent X-ray sources in the field and discussing the data reduction. In Sect.~3 we concentrate on NGC~4410 presenting the HRI results at first before we study the spectral flux distribution of the PSPC data with particular concern of comparison with a sequence of X-ray radiation models. The results are then discussed in Sect.~4 making use of a HST WFPC2 image of NGC~4410a for a geometrical consideration of its inclination.

%__________________________________________________________________

\section{Observations and data reduction}

\begin{figure}
\psfig{figure=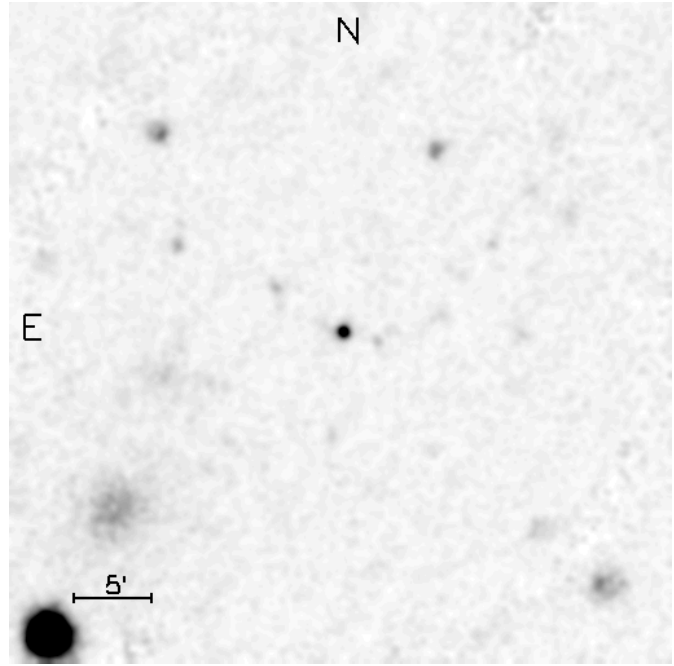,width=8.8cm}
\caption{Central 42\arcmin\ of the PSPC field of view centered around NGC~4410 (bright central source) with an exposure time of 23.4 ksec. The image is smoothed with a Gaussian filter (20\arcsec\ FWHM). The scale of the image is slightly different from the one in Fig. 2. North is up, east is to the left.}
\end{figure}

The data presented in this paper were taken with the HRI and the PSPC detectors on board of the X-ray satellite ROSAT. This X-ray telescope is operating in the energy range of 0.1 to 2.4 keV. The spatial resolutions of the HRI is 1\farcs7. The point spread function (PSF) of this detector at the optical axis in combination with the telescope is 3\arcsec. The PSPC has a PSF of 25\arcsec. The two detectors have a field of view of 38\arcmin\ and 2\degr, respectively. For details concerning ROSAT and its instruments see the ROSAT User's Handbook (Briel et al. 1996).

NGC~4410 was observed on June 28-30, 1993 with the ROSAT PSPC detector for a total effective exposure time of 23.4 ksec. The total number of background-subtracted counts from the central source associated with NGC~4410 is 870$\pm$31. Figure~1 shows the central 42\arcmin\ of the PSPC field of view. For the spectral analysis of the source we used the software package IDL. The source photons were extracted from a circular area of 92\arcsec\ around the central source and corrected for telescope vignetting and detector dead-time. The spectrum of the source was binned according to a signal-to-noise ratio of 8. For the background correction we selected three uncontaminated circular areas close to NGC~4410  with radii of 86\arcsec, 101\arcsec\ and 141\arcsec. The background contributes an X-ray flux of (1.36$\pm$0.27)\tento{-2} cts arcsec$^{-2}$ in the ROSAT bandpass. In order to fit a model spectrum to the PSPC data we used the X-ray spectral-fitting software package XSPEC (Arnaud 1996). 

Our ROSAT HRI observations have been taken between June 27 and July 1, 1995 with a total effective exposure time of 36.7 ksec which led to 152$\pm$24 counts for the central source associated with NGC~4410. The background X-ray flux amouts to (1.6$\pm$1.1)\tento{-2} cts arcsec$^{-2}$.

\subsection{Source detection}

\begin{figure}
\psfig{figure=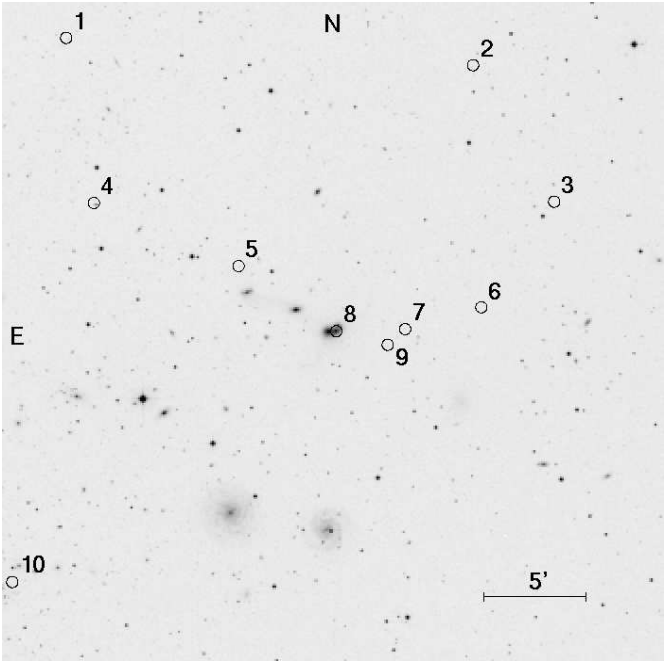,width=8.8cm}
\vspace{-4.2cm}
\caption{ROSAT HRI X-ray sources around NGC~4410 (source no.~8) detected by the maximum likelihood method and with a threshold of 5$\sigma$ ($\sigma$ = 3.0\tento{-7} cts s$^{-1}$ arcsec$^{-2}$), plotted over an optical image taken from the Digitized Palomar Observatory Sky Survey. The size of the field corresponds to the HRI field of view. (Note the different scale compared to Fig. 1.) North is up, east is to the left.}
\end{figure}

Ten X-ray sources above a limit of 5$\sigma$ are found within the field of view of the HRI detector using a standard source-detection algorithm in the EXSAS X-ray analysis software package (Zimmermann et al. 1997) (Fig.~2). In Table~1 the detected sources are listed with position, background subtracted count rates and optical identifications.

\begin{table*}
\caption{ROSAT HRI detections in the NGC~4410 frame with a maximum likelihood threshold of 5$\sigma$.}
\begin{tabular}{llllll}
\hline
No. & RA (2000) & Dec (2000) & countrate (HRI)$^\dag$ & angular distance from & identification\\
& & & [10$^{-3}$ cts~s$^{-1}$] & NGC~4410a$^\ddag$ [arcmin] & \\ \hline \hline
1 & 12$^\mathrm{h}$ 27$^\mathrm{m}$ 16\fs5 & +9\degr 14\arcmin 11\farcs8 & 3.90$\pm$0.50 & 17.70 & QSO 1224+095\\
2 & 12$^\mathrm{h}$ 26$^\mathrm{m}$ 03\fs8 & +9\degr 12\arcmin 59\farcs4 & 2.38$\pm$0.33 & 13.27 &\\
3 & 12$^\mathrm{h}$ 25$^\mathrm{m}$ 49\fs4 & +9\degr 06\arcmin 56\farcs8 & 1.17$\pm$0.25 & 11.30 &\\
4 & 12$^\mathrm{h}$ 27$^\mathrm{m}$ 11\fs5 & +9\degr 06\arcmin 53\farcs3 & 1.97$\pm$0.31 & 12.20 & unidentified optical point source\\
5 & 12$^\mathrm{h}$ 26$^\mathrm{m}$ 45\fs7 & +9\degr 04\arcmin 05\farcs9 & 0.57$\pm$0.15 & 5.21 &\\
6 & 12$^\mathrm{h}$ 26$^\mathrm{m}$ 02\fs4 & +9\degr 02\arcmin 16\farcs4 & 0.90$\pm$0.18 & 6.58 &\\
7 & 12$^\mathrm{h}$ 26$^\mathrm{m}$ 16\fs0 & +9\degr 01\arcmin 18\farcs4 & 0.54$\pm$0.14 & 3.09 &\\
8 & 12$^\mathrm{h}$ 26$^\mathrm{m}$ 28\fs2 & +9\degr 01\arcmin 13\farcs0 & 4.13$\pm$0.36 & 0.00 & NGC~4410a \\
9 & 12$^\mathrm{h}$ 26$^\mathrm{m}$ 19\fs1 & +9\degr 00\arcmin 36\farcs9 & 1.32$\pm$0.20 & 2.39 &\\
10 & 12$^\mathrm{h}$ 27$^\mathrm{m}$ 26\fs0 & +8\degr 50\arcmin 06\farcs7 & 5.30$\pm$0.78 & 18.20 & Abell 1541 \\ \hline
\end{tabular}
\vspace{2mm}
\\
$^\dag$ background subtracted\\
$^\ddag$ NGC~4410a coorinates see no. 8\\
\\
\end{table*}

From these detected X-ray sources in the HRI image, four coincidences with optical objects can be found (see Table~1 and Fig.~2). The X-ray source no.~8 can be clearly identified with NGC~4410a, while an extended X-ray source (no.~10, 44\arcsec\ FWHM) belongs to the galaxy cluster Abell 1541. In addition, the source no.~1 coincides with the QSO 1224+095 and no.~4 is still unidentified in the optical. In order to investigate further identifications of X-ray sources with optical objects we decrease the detection limit to 3$\sigma$. This leads to 32 sources in the HRI image with 10 counterparts in the optical. Three of them are identified as the Abell galaxy cluster 1541 (extended), the QSO 1224+095 (1\arcsec\ offset) and the K7-star G60-2 (8\arcsec\ offset). The offsets between optical and X-ray sources do not indicate any systematical shift or rotation, requiring a position correction of the superposition of the two images at this point.

%__________________________________________________________________

\section{Results}

\subsection{X-ray imaging}

Using the coordinates of the ROSAT HRI pointing the overlay of X-ray and optical image reveals a shift between the optical (NGC~4410a) and X-ray maximum of 4\arcsec. Strikingly, an additional X-ray source, that is visible at the 3$\sigma$ contour level at a distance of $\sim$18\arcsec\ from the central X-ray maximum to the east, has the same displacement from the maximum of the eastern optical component NGC~4410b. It contains about 1\% of the count rate of the central source in the HRI image. From the detection limit of 3$\sigma$ = 9.0\tento{-7} cts s$^{-1}$ arcsec$^{-2}$ in the HRI image and the energy conversion factor (ECF) of 6.7\tento{9} cts cm$^2$ erg$^{-1}$ for the HRI and a power-law spectrum with $\Gamma$ = 2.4 from the ROSAT User's Handbook (Briel et al. 1996) we get an upper limit of 3.8\tento{39} erg s$^{-1}$ for the X-ray luminosity of NGC~4410b.

Furthermore, radio observations of NGC~4410 present a similar emission feature as the X-ray contours. The maximum radio contours correspond spatially to the optical maximum of NGC~4410a and the lower emission levels are elongated toward the southeast (HKG86). To check whether the eastern X-ray maximum has to coincide with the optical structure of NGC~4410, we compare its X-ray luminosity with the blue luminosity $L_{\mathrm B}$ of NGC~4410b. With m$_{\mathrm B}$= 15.28 (MB93) we derive a blue luminosity $L_{\mathrm B}$ = 1.2\tento{10} $L_{\sun}$ for NGC~4410b. Brown \& Bregman (1998) analysed a sample of 34 early-type galaxies observed with ROSAT PSPC and HRI and studied the X-ray-to-optical distribution. The estimated X-ray luminosity of $\sim$ 10$^{39}$ erg s$^{-1}$ for NGC~4410b fits well into the observed distribution of the sample. This obtrudes to superpose the X-ray maxima of both components NGC~4410a and b to the optical image of the pair. This fine tuning requirement is not detectable and therefore not necessary on the scale of the full HRI image (Fig.~2) as pointed out in Sect.~2.1.

\begin{figure}
\psfig{figure=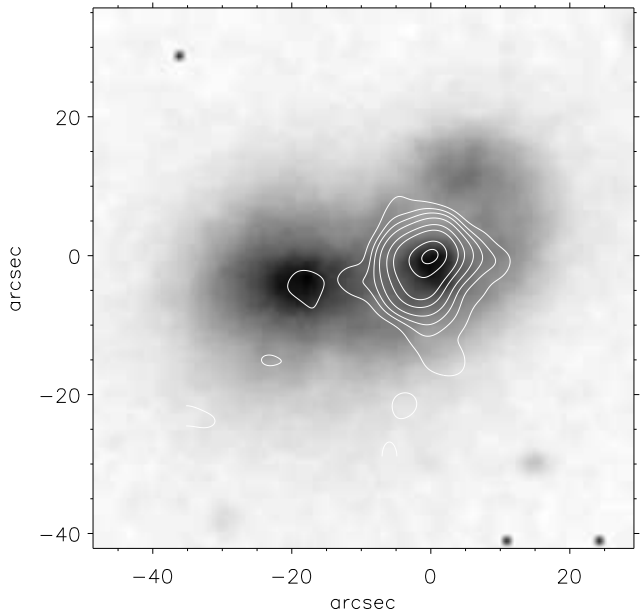,width=12.5cm,angle=90}
\caption{Overlay of the HRI X-ray contours onto the optical image of NGC~4410 taken from the Digitized Palomar sky survey. The X-ray image was smoothed with a Gaussian filter FWHM of 3\arcsec. The contour levels are at 3, 5, 7, 10, 15, 20, 30 and 35$\sigma$ above the mean background with $\sigma$ = 3.0\tento{-7} cts s$^{-1}$ arcsec$^{-2}$. The relative position of optical image and X-ray contours was corrected as explained in Sect. 3.1. North is up, east is to the left.}
\end{figure}

Figure~3 shows the position corrected contour plot of the central 2\arcmin\ of the HRI image with the X-ray source overlaid onto an optical image of NGC~4410 of the Digitized Palomar sky survey plates. The HRI image was smoothed with a Gaussian filter FWHM of 3\arcsec. 

The contours of the HRI image indicate a radial extension from the maximum of NGC~4410a toward the southeast by 3\arcsec\ to 10\arcsec. Since this elongation is close to the resolution limit it has to be checked whether it is real or artificial. Morse (1994) found out that in some ROSAT HRI observations the contours of point-like X-ray sources are elongated over a scale of $\sim$5\arcsec\ to 10\arcsec\ because of errors in the attitude correction as the satellite was wobbled during the observation. On the other hand, it is striking that the extended radio emission of NGC~4410a (HKG86) agrees in its direction with the X-ray contours. Moreover, the elongation of the X-ray contours is not symmetrical with respect to the elongation axis, as one would expect if it is caused by the wobbling satellite, similar to the point-like sources in the paper by Morse (1994). Another possible error namely that the elongation is caused by an incorrect superposition of the different used observation intervalls (OBIs) was also checked by us but didn't explain the effect.

We fit a Gaussian distribution with FWHM 3\farcs5 (spatial resolution of HRI = 1\farcs7, image smoothed with Gaussian filter FWHM 3\arcsec\ corresponding to the HRI PSF) to the point-like source and subtract it from the original image. The result is plotted in the Figs.~4 and 5. The residual reveals a $\sim$10\arcsec$\times$5\arcsec\ emission feature 7\arcsec\ to the southeast of the X-ray maximum visible up to 13$\sigma$ above the background. As expected, the wings of the Gaussian distribution in the Figs. 5b-f fit very well to the X-ray source except in the direction of deformation of the X-ray contours. 

\begin{figure}
\psfig{figure=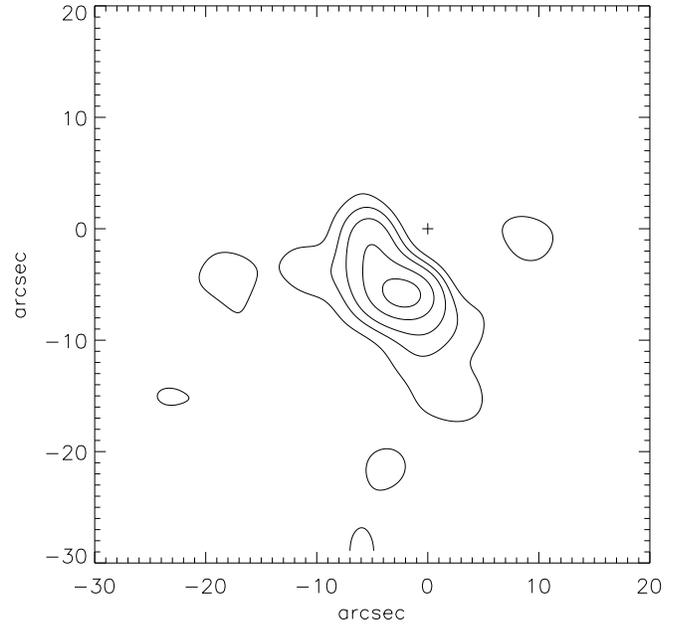,width=9cm,angle=0}
\vspace{-3.6cm}
\caption{Residual X-ray flux after subtracting a point-like source with Gaussian distribution (FWHM 3.5\arcsec) at the position of the maximum of the X-ray flux from the original total flux image (see Fig.~5). The cross marks the position of the maximum and coincides with the nucleus of NGC~4410a. The contours are 3, 5, 7, 10 and 13$\sigma$ above the mean background with $\sigma$ = 3.0\tento{-7} cts s$^{-1}$ arcsec$^{-2}$. North is up, east is to the left.}
\end{figure}

\begin{figure*}
\psfig{figure=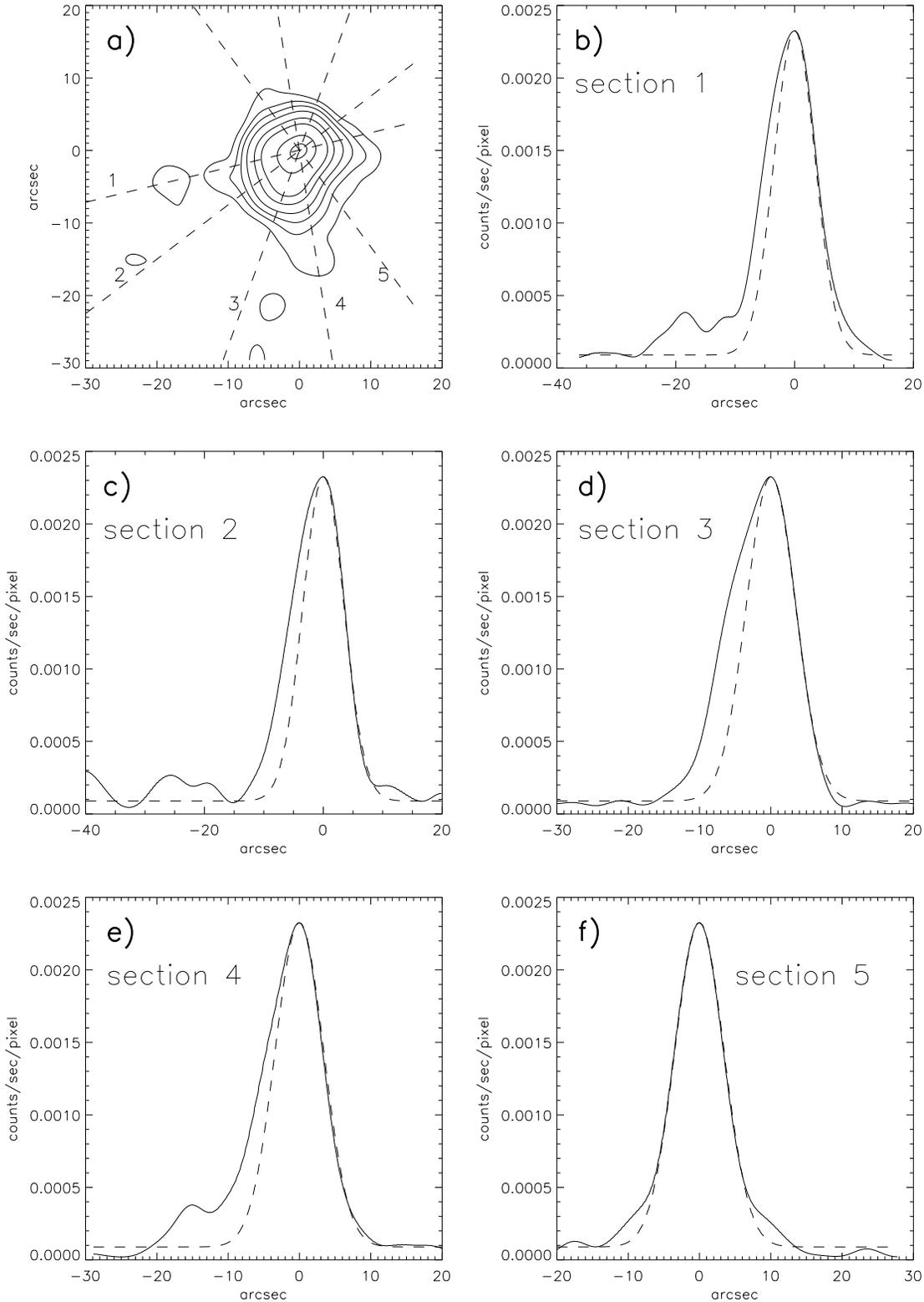,width=15cm,angle=0}
\caption{{\bf a)} Positions of the five sections (dashed lines) through the HRI image of NGC~4410. The contours are the same as in Fig.~3. North is up, east to the left. The following five plots show the corresponding sections through the X-ray maximum of the HRI image. The dashed lines in all five section plots represent the fitted Gaussian distribution with a FWHM 3\farcs5 to the point-like source of NGC~4410a. {\bf b)} Section 1, east-to-west. The local maximum 18\arcsec\ east of the X-ray peak spatially corresponds to NGC~4410b. {\bf c)} Section 2, southeast-to-northwest. {\bf d)} Section 3, southeast-to-northwest. {\bf e)} Section 4, southwest-to-northeast. {\bf f)} Section 5, southwest-to-northeast. In each plot the Gaussian distributed compact source (dashed lines) fits well to the HRI contours except in the direction of deformation of the X-ray contours.}
\end{figure*}

\begin{figure*}
\hspace{1.2cm}
\psfig{figure=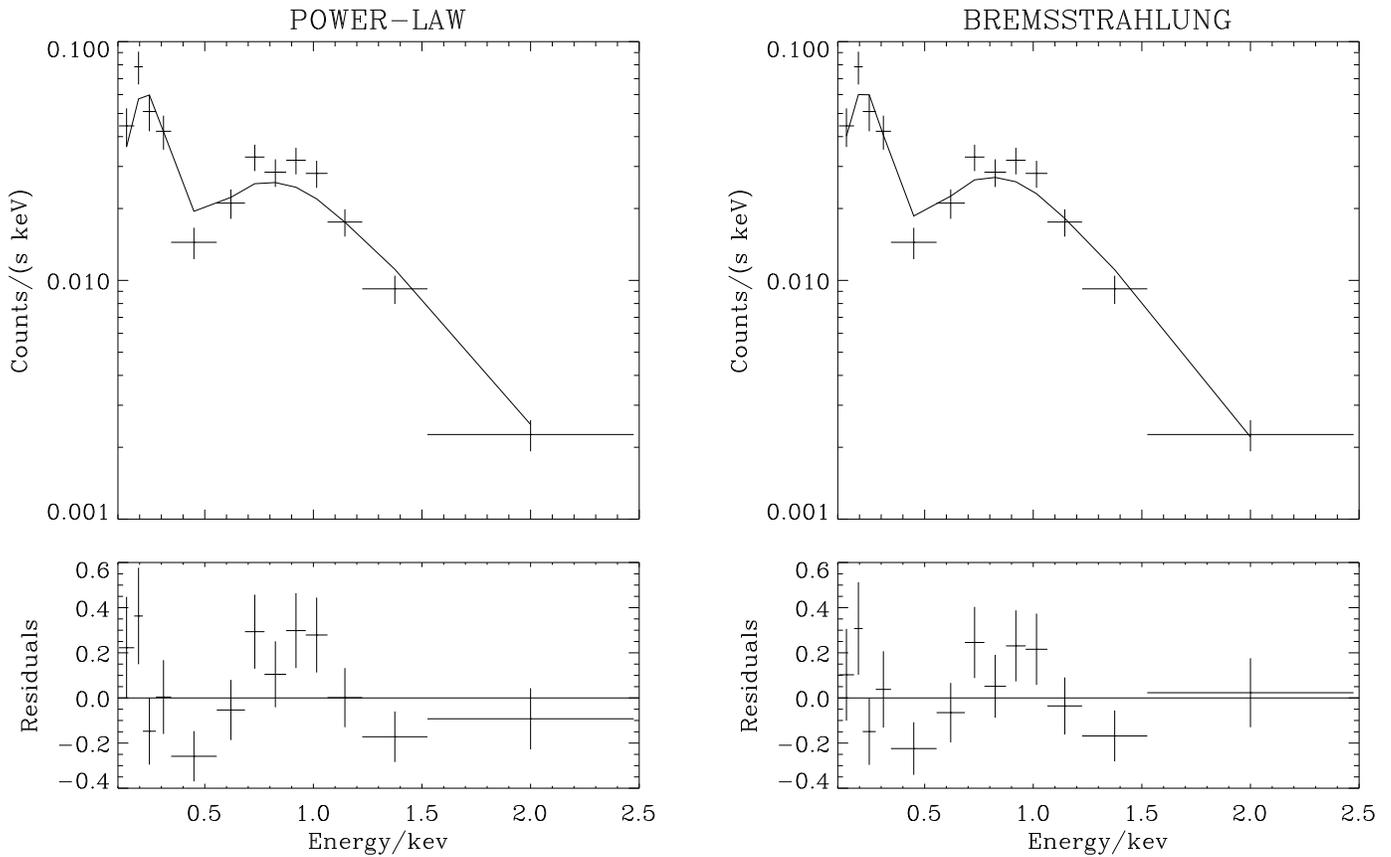,width=16.5cm,angle=90}
\caption{Best single-component model fits to the observed (0.1--2.4) keV X-ray spectrum of NGC~4410, PO (left) and BS (right). The channel counts are binned to a signal-to-noise ratio of 8. The parameter values for these models are shown in Table 2. The lower graphs show the residuals of the fits.}
\end{figure*}

As a working hypothesis which will be supported by the finding of a thermal emission (see Sect.~3.2) we assume that the extended X-ray contours represent a superbubble originating from type {\sc ii} SNe in the starburst region, and expanding into the halo. Since its elongation is too small to resemble the edge-on view of an outflow we further assume that the residual emission in Fig.~5 represents the front part of a bipolar outflow from the central region of the galactic disk. To verify this assumption, we estimate the ratio of the integrated flux from the residual image with the total flux of the original image. For that we integrate the flux within a radius of 13\arcsec\ from the maximum in both the residual image and the original one. We derive a ratio of 0.34$\pm$0.05 (52$\pm$8 counts). The error is estimated by varying the integration radius around the source by 2\arcsec\ and from the mean fluctuation of the background flux. This result is supported by the spectral analysis of PSPC data where a flux of $L_{\mathrm X}$ = 1.3\tento{41} erg s$^{-1}$ for the thermal emission component is derived as part of the combined thermal and power-law spectrum (see~Sect.~3.2). However, since the fraction of the superbubble within the point-like source is neglected, this estimate is only a lower limit.

An X-ray ridge toward the optical position of NGC4410b is discernable at the 3$\sigma$ level, which corresponds to an upper count level of 3.0\tento{-7} cts s$^{-1}$ arcsec$^{-2}$. Using the HRI ECF of 1.8\tento{12} cts cm$^2$ erg$^{-1}$ for a 1 keV Raymond-Smith plasma with log $N_{\mathrm H}$ = 20.0 from the ROSAT User's Handbook (Briel et al. 1996) one gets an upper limit for the X-ray luminosity of 1.4\tento{39} erg s$^{-1}$.
Two regions of enhanced X-ray emission are visible 15\arcsec\ and 22\arcsec\ south of the maximum X-ray source that were not contained in our source list  with a 3$\sigma$ confidence level. Depending on the spectral model and therefore on the ECF the upper limit for the X-ray flux of these sources is 2.25\tento{-16}/ECF erg s$^{-1}$ cm$^{-2}$.

%__________________________________________________________________

\subsection{Spectral analysis}

\begin{table*}
\caption{Spectral fits to the PSPC data of NGC~4410 and the derived model parameters. For the determination of the X-ray luminosity a distance of 97 Mpc is assumed.}
\begin{tabular}{lllllllll}
\hline
model & $N_\mathrm{H}$ & $T$ & $\Gamma$ & norm & red. $\chi^2$ & d.o.f. & $F_\mathrm{X}$ & $L_\mathrm{X}$\\
(1) & (2) & (3) & (4) & (5) & (6) & (7) & (8) & (9)\\ [1.5mm]\hline
\hline
BS & 2.06$^{+0.30}_{-0.27}$ & 11.0$^{+1.6}_{-1.4}$ & -- & 8.0 & 1.6 & 10 & 3.62 & 4.08\\[2mm]
RS (sol) & 1.4$^{+1.2}_{-1.0}$ & 12.7$^{+0.5}_{-0.6}$ & -- & 162 & 4.7 & 10 & 2.96 & 3.33\\[2mm]
RS (sol) & 1.73 $^\dag$ & 13.0$^{+0.5}_{-0.3}$ & -- & 180 & 10.7 & 11 & 2.66 & 2.99\\[2mm]
RS (ww) & (0.06$^{+2000}_{-0.06}$)\tento{-5} & 11.7$^{+0.6}_{-0.8}$ &  -- & 17 & 7.0 & 10 & 3.28 & 3.69\\[2mm]
RS (ww) & 1.73 $^\dag$ & 11.4$\pm$0.3 & -- & 17 & 12.7 & 11 & 2.81 & 3.16\\[2mm]
RS+RS (sol) & 0.45$^{+0.24}_{-0.21}$ & 4.4$^{+1.6}_{-0.8}$ & -- & 35 & 1.0 & 8 & 3.33 & 3.75\\
& & 32$^{+68}_{-13}$ & -- & 224 & & & &\\[2mm]
RS+RS (sol) & 1.73 $^\dag$ & 1.7$^{+0.2}_{-1.0}$ & -- & 100 & 6.6 & 9 & 2.73 & 3.07\\
& & 8.8$^{+3.4}_{-0.5}$ & -- & 65 & & & &\\[2mm]
RS+RS (ww) & 0.35$^{+0.58}_{-0.24}$ & 2.3$^{+0.1}_{-0.2}$ & -- & 2.4 & 0.9 & 8 & 3.65 & 4.10\\
& & 12.0$^{+0.8}_{-1.0}$ & -- & 15 & & & &\\[2mm]
RS+RS (ww) & 1.73 $^\dag$ & 0.9$^{+0.8}_{-0.9}$ & -- & 2.6 & 3.1 & 9 & 2.78 & 3.13\\
& & 9.3$^{+2.0}_{-0.1}$ & -- & 11 & & & &\\[2mm]
RS+PO (sol) & 1.16$^{+0.51}_{-0.33}$ & 7.2$^{+2.3}_{-2.6}$ & 1.85$\pm$0.14 & 76 & 0.8 & 8 & 3.46 (0.88) & 3.90 (0.99)\\[2mm]
RS+PO (sol) & 1.73 $^\dag$ & 7.7$^{+1.6}_{-2.2}$ & 2.08$^{+0.10}_{-0.11}$ & 25 (RS); 8.0 (PO) & 0.7 & 9 & 3.44 (0.78) & 3.87 (0.87)\\[2mm]
RS+PO (ww) & 1.41$^{+0.52}_{-0.25}$ & 9.1$^{+1.4}_{-2.6}$ & 2.03$^{+0.21}_{-0.19}$ & 64 (RS); 4.8 (PO) & 0.7 & 8 & 3.41 (1.15) & 3.84 (1.30)\\[2mm]
RS+PO (ww) & 1.73 $^\dag$ & 9.3$^{+0.9}_{-1.0}$ & 2.17$\pm$0.10 & 5.4 (RS); 6.5 (PO) & 0.6 & 9 & 3.47 (1.20) & 3.91 (1.35)\\[2mm]
RS+PO (ww) $^\#$ & 1.73 $^\dag$ & 1.0$^{+0.3}_{-1.0}$ & & 2.6 (RS) & 1.3 & 8 & 3.20 (0.38) & 3.60 (0.43)\\
& 30.6$^{+5.8}_{-4.4}$ & & 4.24$^{+0.27}_{-0.28}$ & 275 (PO) & & & &\\[2mm]
PO & 3.42$^{+0.30}_{-0.35}$ & -- & 2.39$\pm$0.9 & 12 & 2.3 & 10 & 3.83 & 4.31\\[1mm]
\hline
\end{tabular}
\vspace{2mm}
\\
$^\dag$ fixed to Galactic foreground value\\
$^\#$  separate absorption for each component\\
\\
Col.(1)--- Emission models abbreviated as: BS = thermal Bremsstrahlung, RS = Raymond-Smith, PO = power-law. The specification in brackets indicates the used element abundances for the hot thermal plasma, solar (sol) or abundances of a supernova calculated by Woosley and Weaver (ww) and integrated over all stars between 10 and 100 M$_{\sun}$ for a Salpeter IMF.\\ 
Col. (2)--- column density in units of 10$^{20}$ cm$^{-2}$.\\ 
Col. (3)--- plasma temperature in units of 10$^6$ K.\\ 
Col. (4)--- photon index.\\ 
Col. (5)--- scaling factor: for BS in units of (10$^{-19}$/(4$\pi D^2$))$\int n_\mathrm{e}n_\mathrm{I}$d$V$, $n_\mathrm{e}, n_\mathrm{H}$ = electron and ion densities (cm$^{-3}$); for RS in units of (10$^{-20}$/(4$\pi D^2$))$\int n_\mathrm{e}n_\mathrm{H}$d$V$, $n_\mathrm{e}, n_\mathrm{H}$ = electron and H densities (cm$^{-3}$); for PO in units of 10$^{-5}$ photons~keV$^{-1}$ cm$^{-2}$ s$^{-1}$ at 1 keV.\\ 
Col. (6)--- reduced $\chi^2$.\\ 
Col. (7)--- degrees of freedom.\\ 
Col. (8)--- X-ray flux in units of 10$^{-13}$ erg cm$^{-2}$ s$^{-1}$, corrected for the fitted absorption. Values in brackets give the contribution of the thermal component.\\ 
Col. (9)--- X-ray luminosity in units of 10$^{41}$ erg s$^{-1}$. Values in brackets give the contribution of the thermal component.
\end{table*}

\begin{figure}
\psfig{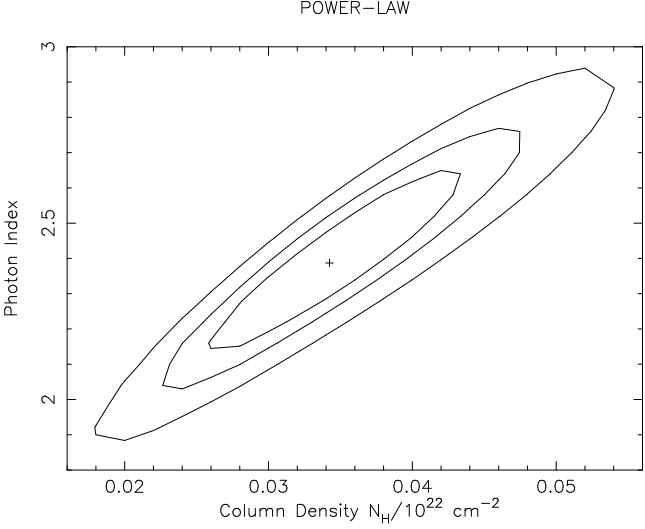}
\caption{$\chi^2$ contours of the fit parameters column density $N_\mathrm{H}$ and photon index $\Gamma$ of the PO model (Fig.~6). The levels represent 99\%, 90\% and 68\% confidence.}
\end{figure}

\begin{figure}
\psfig{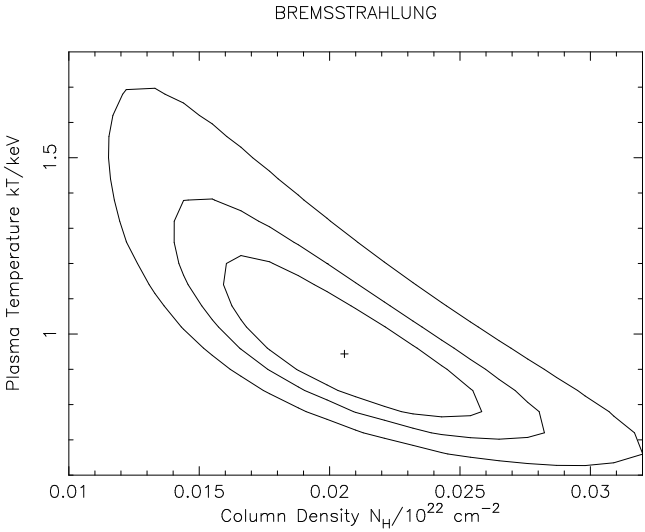}
\caption{$\chi^2$ contours of the fit parameters column density $N_\mathrm{H}$ and plasma temperature $kT$ of the BS model (Fig.~6). The levels represent 99\%, 90\% and 68\% confidence.}
\end{figure}

\begin{figure}
\psfig{figure=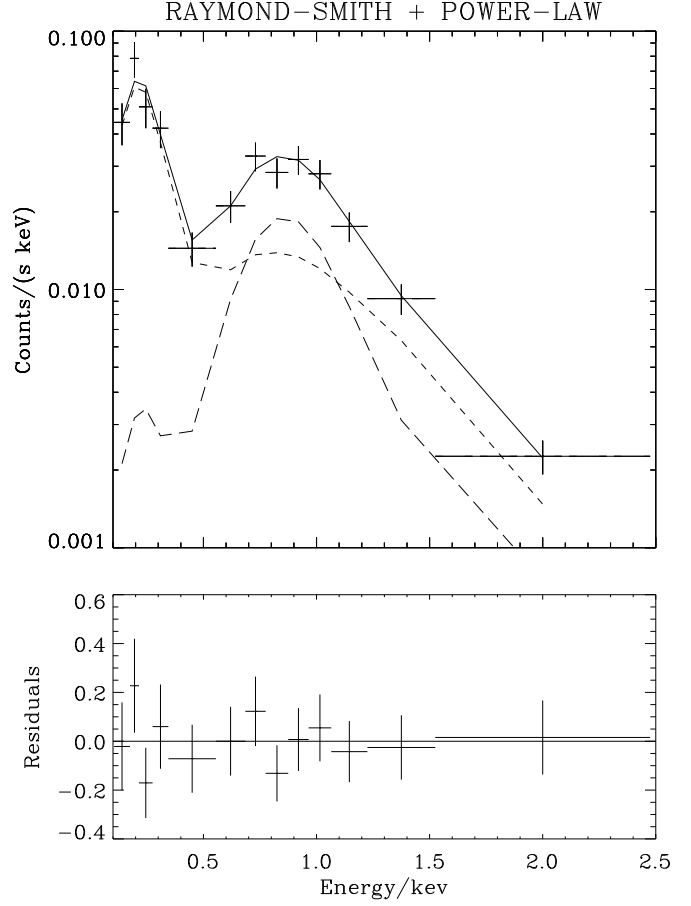,width=8.8cm}
\caption{Observed X-ray spectrum of NGC~4410 and best fit (solid line) with a combined RS and PO model(RS+PO(ww)). The parameter values for this model are shown in Table 2. The long-dashed line shows the spectrum of the thermal component. The short-dashed line shows the contribution of the power-law. The contribution of the RS component to the total flux is 35\%. The residuals of this fit are plotted in the lower box.}
\end{figure}

\begin{figure}
\psfig{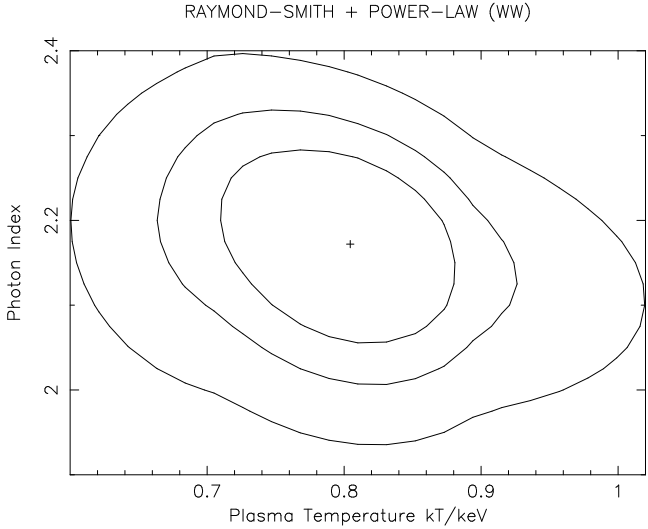}
\caption{$\chi^2$ contours of the fit parameters plasma temperature $kT$ and photon index $\Gamma$ of the two-component RS+PO(ww) model (Fig.~9). The levels represent 99\%, 90\% and 68\% confidence.}
\end{figure}

\begin{figure}
\psfig{figure=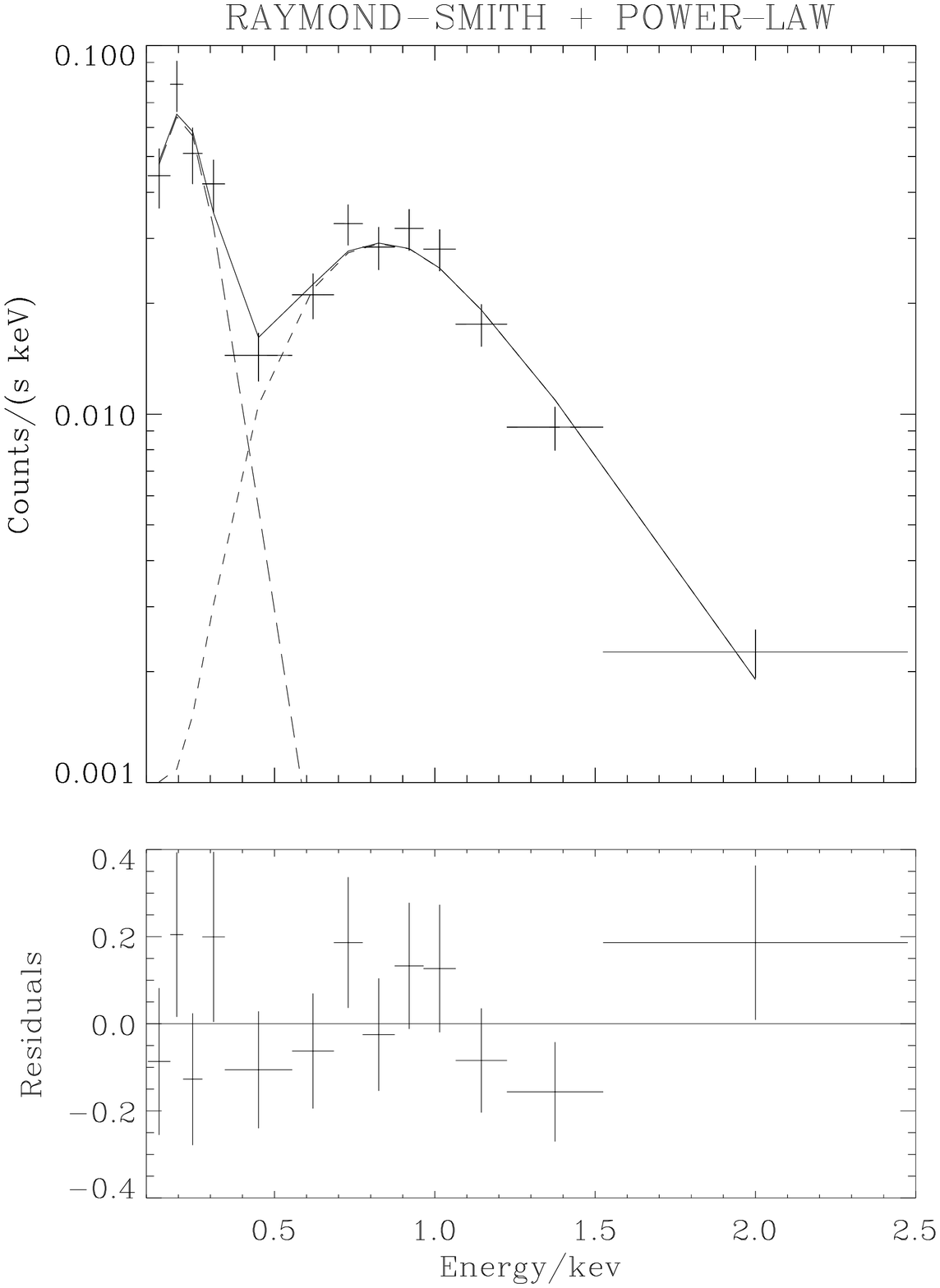,width=8.8cm}
\caption{Fit of the RS+PO(ww) model with individual column densities for each component and residuals of the fit. The different lines represent the same components as in Fig.~9. As expected the soft part of the spectrum is dominated by the unabsorbed thermal emission from the hot plasma. The hard spectrum $>$0.5 keV originates from the power-law component. Yet the spectrum can only be fitted with much too steep photon index of more than 4.}
\end{figure}

In addition to the spatial information of emission components we got from the HRI data, we now want to decompose these components by using the spectral information obtained from the PSPC data. With the PSF of 25\arcsec\ of the PSPC we get no satisfactory spatial resolution.

The 870 background corrected counts of the central source detected with the PSPC in the energy range of 0.1-2.4 keV were binned in order to achieve a constant signal-to-noise ratio of 8. To fit the observed PSPC X-ray spectrum we apply different models of absorbed emission spectra including a thermal Bremsstrahlung spectrum (BS), the spectrum of an optically thin thermal plasma in collisional equilibrium (RS) as calculated by Raymond \& Smith (1977) and a power-law spectrum (PO) with a photon energy distribution $\propto E^{-\Gamma}$, as well as several combinations of those. The absorption of the spectra depends on the hydrogen column density $N_\mathrm{H}$ between the source and the observer. As a first-order estimate and as a lower limit of $N_\mathrm{H}$ we use the Galactic foreground value in the direction of NGC~4410. This value for $N_\mathrm{H}$ amounts to 1.73\tento{20} cm$^{-2}$ (Dickey \& Lockman 1990).

The model fits for the X-ray emission to the observation are shown in Table~2. Usually the RS model is applied under the assumption of solar abundances (Anders \& Grevesse 1989). Reasonably, this is not an ideal approach because expanding SN ejecta should contain the nucleosynthesis products of their massive stellar progenitors. For this reason we use two different sets of chemical abundances. Beside the classical model with solar abundances we fit one with enhanced abundances of SNe according to the yields by Woosley \& Weaver (1985), but integrated over all massive stars between 10 and 100 M$_{\sun}$ and with a Salpeter initial mass function (IMF).

The main uncertainty to fit the data comes in by the value of the hydrogen column density $N_\mathrm{H}$. In each model, except the combinations of RS+PO and the single component models BS and PO, the free parameter $N_\mathrm{H}$ turns out to become much too low for the best fit. For the single thermal model RS it is not possible to fit both, the soft and the hard part of the spectrum reasonably without an unrealistically low value of $N_\mathrm{H}$. 

This is also true for the combined two temperature plasma model RS+RS(ww). The best fit results in an unrealistically low $N_\mathrm{H}$ value of 0.35\tento{20} cm$^{-2}$. For the same model with solar abundances RS+RS(sol), the best fit yields $N_\mathrm{H}$ = 0.45\tento{20} cm$^{-2}$. In addition, the plasma temperature of more than 3\tento{7} K for the second component is abnormally high for halo gas. This temperature may only be reached for gas in the central regions of starburst galaxies, for example NGC 253 (Ptak et al. 1997) and M82 (Mathews \& Doane 1994). In the case of fixing the column density to the Galactic value, the resulting plasma temperatures of 1.7\tento{6} K and 8.8\tento{6} K for RS+RS(sol) and 0.9\tento{6} K and 9.3\tento{6} K for RS+RS(ww) are much more reasonable. But both fits to the observed spectra are much worse (see Table~2).

In the case of fitting the observation with a single PO the column density of 3.4\tento{20} cm$^{-2}$ becomes plausible because of an expected additional absorption within NGC~4410. The photon index of 2.39$\pm$0.09 is in agreement with the results from Turner et al. (1993) and Mulchaey et al. (1993), who found a mean value for Seyfert\,1 and Seyfert\,2 galaxies of $\Gamma$ = 2.4. Comparing all these single-component models the model PO and BS provide the best fits to the observation. The fits are plotted in Fig.~6.

The best fit by far for the whole observed spectrum can be reached by applying a two-component model which combines a thermal and a power-law spectrum (RS+PO) with a photon index of 2.17$\pm$0.10 and a plasma temperature of (9.3$\pm$1.0)\tento{6} K (see Fig.~9). Nevertheless, the fit becomes worse for applying a higher value for $N_\mathrm{H}$ than the Galactic one. The combined PO+RS fit yields the best result with an absorption corrected luminosity of (3.91$\pm$0.55)\tento{41} erg~s$^{-1}$ for the assumed distance of 97~Mpc.

Under the assumption, that the nonthermal emission originates from a central active nucleus heavily obscured by dust and gas, and that hot plasma has expanded into the outer parts of NGC~4410a, we fit the RS+PO model with different column density values for each component. One would expect a column density $N_\mathrm{H,RS} \approx N_\mathrm{H,Gal}$ for the less obscured plasma outflow and a much higher $N_\mathrm{H,PO}$ for the AGN because of the intrinsic absorption within the NGC~4410 nucleus. While we apply $N_\mathrm{H,RS}$ = $N_\mathrm{H,Gal}$ = 1.73\tento{20} cm$^{-2}$ for the RS component, $N_\mathrm{H,PO}$ for the PO component is set to 3.1\tento{21} cm$^{-2}$. This yields the best fit for a lower plasma temperature of 1.0\tento{6} K and a very steep power-law for the central AGN with $\Gamma$ =  4.24 (see~Fig.~11). In the very soft range (0.1-0.3 keV) the spectrum is determined by the intrisically unobscured RS component. Raising the absorption for the PO component prohibits its contribution to the soft spectral range, which means that the total X-ray flux there has to originate from the hot ionized gas. As a consequence the temperature of the plasma is almost one order of magnitude lower than in the RS+PO model with only one common absorption. The hard spectrum $>$0.5 keV originates completely from the nonthermal component. Nevertheless the photon index is unplausibly steep.

%__________________________________________________________________

\section{Discussion}

\subsection{The X-ray halo}

The HRI image strongly suggests a twofold spatial correlation between the X-ray and the radio emission: we have a compact central emission peak apparently coinciding with NGC~4410a and a diffuse halo slightly elongated about 10\arcsec\ toward the southeast. Subtraction of the central point-like source from the total X-ray emission enables us to unveil a clear extention in this direction. Because of the asymmetrical elongation and the spatial correlation in the X-ray and radio flux we are convinced not to be mistaken by an artefact. Furthermore, the flux ratio of 0.34$\pm$0.05 from this feature compared to the total flux corresponds to the best fitting model for the observed spectrum, a two-component model, combining a thermal Raymond-Smith spectrum and a power-law spectrum. As a hypothesis we assume that the thermal emission originates from an expanding X-ray halo gas with a plasma temperature of a few 10$^{6}$ K whose expulsion is driven by a central engine (SB + AGN).

\begin{figure*}
%\psfig{figure=fig12.ps,width=18cm}
%\psfig{figure=, width=18cm}
%\picplace{4cm}
\vspace{18.0cm}
\caption{Broad band image in the V (5843 \AA) filter of the nucleus of NGC~4410a with a logarithmic grayscale taken from the Hubble Space Telescope archive. As a consequence of the interaction, a chain of knots and dust, a bright emission region near the nucleus and a long tidal arm are visible. The image has a size of 34\arcsec$\times$34\arcsec. North is up, east is to the left.}
\end{figure*}

The SE elongation of the emission contours could be caused by two effects:\\
(1) If the pec~Sab galaxy NGC~4410a is seen at moderate inclination, a circumnuclear superbubble or a bipolar outflow is visible in soft X-ray which expands vertically to both sides out of the galactic disk only the part in front of the H{\sc i} disk. X-ray emission in the ROSAT band originating from gas outflowing into the reverse halo hemisphere is absorbed by cold gas in the disk of NGC~4410. This fact also accounts for the low N$_{\mathrm H}$ that is required from the best model fit and is lacking of any additional contribution to the Galactic foreground value.\\
(2) The X-ray halo gas is exposed to the tidal force of the companion galaxy NGC~4410b. 

In order to decide which of the two effects causes the deformation of the X-ray halo, information about the inclination angle of NGC~4410a is necessary. Fig.~12 shows the {\it Hubble Space Telescope} (HST) WFPC2 image of the central part of NGC~4410a in the V (5843 \AA) broad band filter taken from the Hubble Space Telescope archive. One can discern the nucleus, a bright emission region at the lower left, and a line of additional smaller bright knots at the left. At the right-hand side of the nucleus a long tidal arm probably caused by the interaction is clearly visible. Since the morphology in this image leads to a very low inclination, we prefer the second explanation of an interaction-induced deformation of the X-ray halo. However, the HST image suggests an additional possibility, namely, that the bright emission region at the lower left is a starburst region, and that the extended part of the X-ray contours originates from this region.

The X-ray contours reveal a faint ridge toward NGC~4410b ($L_{\mathrm X} \le$ 1.4\tento{39} erg s$^{-1}$) which has only a very faint corresponding X-ray counterpart centrally of NGC~4410b ($L_{\mathrm X} \le$ 3.8\tento{39} erg s$^{-1}$) as indication of any activity in or around the nucleus of NGC~4410b. The non-detection of a similar radio emission admits at best a radioquiet AGN.

\subsection{Physical conditions of the X-ray emission}

As one can see in Table~2 the derived values for the X-ray luminosity are all of the same order of ~3--4\tento{41} erg s$^{-1}$ and independent of the applied model. From this point of view it is not possible to distinguish between the different models. The two most acceptable models with reasonable values for N$_{\mathrm H}$ are the two-component combination RS+PO and the single-component PO model, respectively.

Although the photon index of $\Gamma$ = 2.39$\pm$0.09 of the single PO model is in good agreement with observed values for low luminousity AGNs (Turner et al. 1993; Mulchaey 1993), it is obvious that the observed X-ray spectrum of NGC~4410 cannot be fitted properly by a single PO model but requires an additional component as can be seen in Table~2 and by comparing Fig.~6 with Fig.~9. The RS+PO model for the X-ray source leads to a luminosity of the thermal component $L_\mathrm{X}$ =  (1.35$\pm$0.19)\tento{41} erg s$^{-1}$ (35\% of the total X-ray luminosity). Nevertheless, the X-ray luminosity can vary by a factor of 4 because of the uncertainty of the distance to the source. To compare the derived properties of NGC~4410 with theoretical calculations, we apply the following model: We assume a central SB to drive an expanding superbubble. The central source injects mechanical energy since a time $t_\mathrm{exp}$ at a constant rate $L_\mathrm{mech}$ by sequential supernova explosions into the bubble. The supersonic expansion leads to a shock front and is heating the ambient gas. Assuming a homogeneous density of this ambient gas would lead to a spherically expanding bubble, as calculated by MacLow \& McCray (1988). 

This model, however, does not correspond to the realistic conditions of a SB in a disk galaxy. Because of the vertically decreasing gas density the superbubble forms a "chimney" and blows out of the galactic disk into the halo by a bipolar outflow (Norman \& Ikeuchi 1989). The dynamics, X-ray emission and disk-halo interaction of these outflows are modelled under different conditions of ambient disk gas density, halo gas density, energetic input into the superbubble, and expansion time of the superbubble in a number of papers (Tomisaka \& Ikeuchi 1986; Tomisaka \& Ikeuchi 1988; MacLow et al. 1989; Suchkov et al. 1994, hereafter S94; Michaelis et al. 1996). S94 found that the soft X-ray emission of the suberbubble in the range of (0.1--2.2) keV primarily originates from the shock-heated halo and disk gas with temperatures of 10$^6$ -- 10$^7$ K, rather than from the supernova material itself. Depending on the density of the disk and halo gas, the scale-height of the disk, and the energy deposition rate from the SB, the morphology of the bipolar outflow varies in structural parameters: opening angle within the disk, radius of the chimney, vertical extension of the superbubble, and ratio of the two last properties. Independent of  these different conditions, the models always achieve a bipolar morphology. S94 derived an analytic expression for the shell X-ray luminosity in the (0.1--2.2) keV band:

\begin{equation}
L_{\mathrm{X}} = 9.7\mtento{40} (L_{42})^{3/5} (n_{0.01})^{7/5} (t_{7})^{9/5} \mathrm{erg\ s^{-1}} ,
\end{equation}

\noindent where $L_{42}$ is the mechanical input energy in units of 10$^{42}$ erg s$^{-1}$, $n_{0.01}$ is the particle density of the halo gas in units of 0.01 cm$^{-3}$, and $t_{7}$ is the expansion time of the superbubble in units of 10$^{7}$ yr.

From the HRI observation we can estimate a maximum expansion of the outflowing gas into the halo of ~10\arcsec, scaling to 4.7 kpc at a distance of 97 Mpc. The models of Tomisaka \& Ikeuchi (1986), Tomisaka \& Ikeuchi (1988), MacLow et al. (1989) and S94 reveal chimney radii within the galactic disk between 200 pc and 600 pc. The much more extended X-ray emission observed in NGC~4410a can therefore only originate from the outflow into the less dense galactic halo where it can expand in each direction. This involves that the SB has to be old enough in order to account for the escape of the expanding superbubble from the disk into the halo. That means a lower limit of $t_\mathrm{exp} >$ 4 Myr for the expansion time (Tomisaka \& Ikeuchi 1988).

We estimate the mechanical energy input by the starburst into the superbubble from the emitted H$\alpha$ radiation using the shock model by Binette et al. (1985). They calculated the radiative cooling mechanism of shock-heated gas, emitting optical line radiation, and found that $L_\mathrm{H\alpha} \approx$ 10$^{-2} L_\mathrm{mech}$. With $L_\mathrm{H\alpha}$ = 6.6\tento{40} erg s$^{-1}$ for NGC~4410 (MB93) this leads to $L_\mathrm{mech} \approx$ 6\tento{42} erg s$^{-1}$. Applying 10$^{51}$ erg for the energy release per \SNII and taking into account that roughly only 20\% is converted into mechanical luminosity we derive a \SNII rate of $\sim$1.0 yr$^{-1}$. 

Under the simplified consideration of a spherically expanding gas one can estimate its density. With the scaling factor of the RS model and setting the electron density equal to the hydrogen density we get the following expression for the electron density of the hot gas:

\begin{equation}
n^2_e = \frac{3D^2 N_{RS}}{fr^3}\times10^{14}cm^{-5},
\end{equation}

\noindent where $N_{RS}$ is the scaling factor, $D$ is the distance to the source, $r$ is the radius of the superbubble and $f$ is a filling factor taking into account that the hot gas is not distributed homogeneously but broken up into separate bubbles. With the parameters of $D$ = 97 Mpc, $r$ = 4 kpc, $N_{RS}$ = 5\tento{-6} cm$^5$ and an assumed filling factor of 0.1 one obtains an electron density of 0.03 cm$^{-3}$. Changing the filling factor to $f$ = 0.9 leads to $n_e$ = 0.01 cm$^{-3}$.

An expansion time of 10$^7$ yr with a halo gas density of 0.01 cm$^{-3}$ leads to an X-ray luminosity of the shell of 2.8\tento{41} erg s$^{-1}$. A lower \SNII rate of 0.5 yr$^{-1}$ and a slightly smaller expansion time of 8\tento{6} yr reduce the obtained X-ray luminosity to 1.2\tento{41} erg s$^{-1}$ in the ROSAT band. The derived plasma temperature of 10$^{7}$ K lies at the upper bound of the range with log $T$ = 6.0 -- 6.9 found by S94 for the (0.1--2.2) keV band.

For a Salpeter IMF, a \SNII activity between 10 and 100 M$_{\sun}$ and a SN rate of 0.5 yr$^{-1}$ the star formation rate results to $\sim$95 M$_{\sun}$ yr$^{-1}$. Depending on the fraction of the mechanical energy release of a \SNII this value can increase up to a factor of 5.

\subsection{Comparison with other galaxies of similar properties}

Each galaxy and, in particular, mergers, galaxy pairs or SB galaxies are unique systems. In order to get an insight on whether NGC~4410 and its derived structures are somehow typical for close encounters, we compare the derived X-ray luminosity in the ROSAT band with other disturbed and isolated SB galaxies. The peculiar galaxy NGC~2782 e.g. is thought to be a merger of two disk galaxies of unequal mass and has $L_\mathrm{X}$ = 4\tento{41} erg s$^{-1}$ (Schulz et al. 1998). Another galaxy with disturbed morphology and comparable X-ray luminosity ($L_\mathrm{X}$ = 1.4\tento{41} erg s$^{-1}$) is NGC~1808. In contrast the PSPC data of this object do not show any X-ray outflow out of the central source into the halo. But one has to mention that NGC~1808 has a SFR of only 10 M$_{\sun}$ yr$^{-1}$ (Junkes et al. 1995). Relatively isolated systems without any companion, like e.g. NGC~253 (Fabbiano et al. 1992), NGC~2903 and NGC~4569 (Junkes et al. in preparation), contain X-ray luminosities of a few 10$^{40}$ erg s$^{-1}$, emphasizing the importance of interaction for star-forming activity.

PC96 found a significant difference in the $L_\mathrm{X}/L_\mathrm{H\alpha}$ ratio between pure AGN, pure SBs and galaxies with circumnuclear star-forming rings with an active nucleus. The pure active nuclei show log($L_\mathrm{X}/L_\mathrm{H\alpha}$) between 0.00 and +1.68, while the pure SBs in the sample lie between -1.46 and -0.36. The three galaxies with combined X-ray emission from AGN and SB have values of -0.26 (NGC~1097), +0.16 (NGC~1068) and +0.63 (NGC~7469), indicating a continuous decrease from AGN to SB. From this tendency one would expect a log($L_\mathrm{X}/L_\mathrm{H\alpha}$)$<$0 for the RS+PO model for NGC~4410. Our results, however yields +0.77 thereby, approximately the same as for a single PO model with log($L_\mathrm{X}/L_\mathrm{H\alpha}$) = +0.81. 

The fraction of $L_\mathrm{H\alpha}$ from the SB relative to the total H$\alpha$ luminosity amounts to 98\%, 80\% and 40\% for NGC~1097, NGC~1068 and NGC~7469, respectively. Comparing only the contributions from the SB to the H$\alpha$ and X-ray luminosity, PC96 found log($L_\mathrm{X}/L_\mathrm{H\alpha}$) = -0.99, -0.70 and -0.36 for NGC 1097, NGC~1068 and NGC~7469, respectively. If we concider the fraction of H$\alpha$ luminosity from the SB in these galaxies, and assume that 90\% of the total H$\alpha$ luminosity originates from the SB within NGC~4410a, $L_\mathrm{H\alpha}$ would result in 5.9\tento{40} erg s$^{-1}$ and log($L_\mathrm{X}/L_\mathrm{H\alpha}$) = +0.36, which is quite high compared to the sample analysed by PC96. 

%__________________________________________________________________

\section{Conclusions}

We observed the interacting pair of galaxies NGC~4410 with the ROSAT HRI and PSPC. Spectral investigations of NGC~4410 suggest that the integral X-ray emission ($L_\mathrm{X}$ = 3.9\tento{41} erg s$^{-1}$) can be decomposed into a thermal component (described by a RS spectrum) and a component from the AGN (described by a power-law spectrum). The HRI image reveals an extended X-ray halo related to NGC~4410a with an extension of 10\arcsec\ from the nucleus of NGC~4410a to the southeast. Combining spatial and spectral informations reveals an X-ray luminosity of the halo gas of $L_\mathrm{X}$ = 1.3\tento{41} erg s$^{-1}$ (1/3 of the total X-ray emission). The companion galaxy NGC~4410b houses only a very faint central point-like source below the 3$\sigma$ level, corresponding to an upper limit of 3.8\tento{39} erg s$^{-1}$ for the X-ray luminosity.

As a reasonable model we can assume that the tidal interaction in the pair of galaxies NGC~4410 has two effects on the one partner, the face-on pec~Sab galaxy NGC~4410a:\\
(1) A central monster is either formed due to this interaction or has already existed before and is now fed by infalling gas during the merging event, producing AGN signatures. Evidence for an existing AGN comes from the ROSAT X-ray spectrum supported by the spatially correlated radio emission.\\
(2) The merging process has ignited a nuclear SB ejecting an X-ray gas into the halo. Due to the tidal forces of NGC~4410b, the X-ray halo of NGC~4410a is deformed toward its neighbouring galaxy, either directly by gravitational forces or indirectly, originating from a decentred starburst region as a consequence of the merging process.

This system of merging galaxies represents a highly interesting object because it is a close interacting pair where the effects of tidal forces on nuclear activity and SB can be studied in detail with large telescopes. The HST WFPC2 image of NGC~4410a suggests a possible relation between the deformed X-ray halo and a bright emission region near the nucleus. With ROSAT imaging it is not possible to distinguish between the nucleus and the bright emission region resolved by HST. Further spectroscopic observations of this region and additional infrared imaging could help answering the question of the exact origin of the X-ray emission. 

%__________________________________________________________________

\begin{acknowledgements}
The authors are grateful to an unknown referee for his substantial and constructive report. The ROSAT project is supported by the German Bundesministerium f\"ur Bildung, Wissenschaft, Forschung und Technologie (BMBF) and the Max-Planck-Society. The optical image shown is based on photographic data of the National Geografic Society - Palomar Observatory Sky Survey (NGS-POSS) obtained using the Oschin Telescope on Palomar Mountain. The NGS-POSS was funded by a grant from the National Geographic Society to the California Institute of Technology. The plates were processed into the present compressed digital form with their permission. The Digitized Sky Survey was produced at the Space Telescope Science Institute (STScI) under US Government grant NAG W-2166. This research has made use of the NASA/IPAC Extragalactic Database (NED) which is operated by the Jet Propulsion Laboratory, Caltech, under contract with the NASA. Observations made with the NASA/ESA Hubble Space Telescope were used, obtained from data archive at STScI. STScI is operated by the Association of Universities for Research in Astronomy, Inc. (AURA) under the NASA contract NAS 5-26555.
\end{acknowledgements}

%__________________________________________________________________


\begin{thebibliography}{}
\bibitem[1989]{anders}
Anders E., Grevesse N. 1989, Geochimica et Cosmochimica Acta 53, 197
\bibitem[1996]{arnaud}
Arnaud K.A. 1996, Astronomical Data Analysis Software and Systems V, ASP Conf. Ser. vol. 101, eds. Jacoby G. and Barnes J., p. 17
\bibitem[1992]{batuski}
Batuski D.J., Hanisch R.J., Burns J.O. 1992, AJ 103, 1077
\bibitem[1995]{bicay}
Bicay M.D., Kojoian R.J., Seal J., et al. 1995, ApJS 98, 369
\bibitem[1985]{binette}
Binette L., Dopita M.A., Tuohy I.R. 1985, ApJ 297, 476
\bibitem [1996]{briel}
Briel U., Aschenbach B., Hasinger G., et al. 1996, ROSAT User's Handbook (MPE, Garching)
\bibitem[1998]{brown}
Brown B.A., Bregman J.N. 1998, ApJ 495, L75
\bibitem[1986]{bushouse}
Bushouse H.A. 1986 AJ 91, 255
\bibitem[1990]{dickey}
Dickey J.M., Lockman F.J. 1990, ARAA 28, 215
\bibitem[1992]{fabbiano}
Fabbiano G., Kim D.-W., Trinchieri G. 1992, ApJS 80, 531
\bibitem[1998]{genzel}
Genzel R., Lutz D., Sturm E., et al. 1998, ApJ 498, 579
\bibitem[1986]{hummel}
Hummel E., Kotanyi C.G., van Gorkom J.H. 1986, A\&A 155, 161 (HKG86)
\bibitem[1992]{jog1}
Jog C.J., Das M. 1992, ApJ 400, 476
\bibitem[1992]{jog2}
Jog C.J., Solomon P.M. 1992, ApJ 387, 152
\bibitem[1984]{joseph}
Joseph R.D., Meikle W.P.S., Robertson N.A., Wright G.S. 1984, MNRAS 209, 111
\bibitem[1995]{junkes}
Junkes N., Zinnecker H., Hensler G., et al. 1995, A\&A 294, 8
\bibitem[1985]{keel}
Keel W.C. 1985, in "Astrophysics of Active Galaxies and Quasi-Stellar Objects", ed. Miller J.S., University Science Books, Mill Valley, p.1
\bibitem[1984]{lonsdale}
Lonsdale C.J., Persson S.E., Matthews K. 1984, ApJ 287,95
\bibitem[1998]{lutz}
Lutz. D., Spoon H.W.W., Rigopoulou D., et al. 1998, ApJ 505 L103
\bibitem[1988]{maclow}
MacLow M.-M., McCray R. 1988, ApJ 324, 776
\bibitem[1989]{maclow2}
MacLow M.-M., McCray R., Norman M.L. 1989, ApJ 337, 141
\bibitem[1994]{mathew}
Mathew W.G., Doane J. 1994, in "Panchromatic View of Galaxies -- Their Evolutionary Puzzle", eds. Hensler G., Theis C., Gallagher J.S., \'Editions Fronti\`eres, p. 221
\bibitem[1993]{mazzarella1}
Mazzarella J.M., Boroson T.A. 1993, ApJS 85, 27 (MB93)
\bibitem[1991]{mazzarella2}
Mazzarella J.M., Bothun G.D., Boroson T.A. 1991 AJ 101, 2034
\bibitem[1996]{michaelis}
Michaelis O., Hensler G., Samland M. 1996, in IAP Proc. "The Interplay between Massive Star Formation, the ISM and Galaxy Evolution", eds. Kunth D. et al.,Editions Frontiers, Gif-sur-Yvettes, p. 525
\bibitem[1994]{morse}
Morse J.A. 1994, PASP 106, 675
\bibitem[1993]{mulchaey}
Mulchaey J.S., Colbert E., Wilson A.S., et al. 1993, ApJ 414, 144
\bibitem[1989]{norman1}
Norman C.A., Ikeuchi S. 1989, ApJ 345, 372
\bibitem[1988]{norman2}
Norman C.A., Scoville N.Z. 1988, ApJ 332, 124
\bibitem[1996]{perez}
P\'erez-Olea D.E., Colina L. 1996, ApJ 468, 191 (PC96)
\bibitem[1997]{ptak}
Ptak A., Serlemitsos P., Yaqoob T., Mushotzky R. 1997, AJ 113, 1286
\bibitem[1977]{raymond}
Raymond J.C., Smith B.W. 1977, AJS 35, 419
\bibitem[1984]{rees}
Rees M.J. 1984, ARAA 22, 471
\bibitem[1996]{sanders}
Sanders D.B., Mirabel I.F. 1996, ARAA 34, 749
\bibitem[1998]{schulz}
Schulz H., Komossa S., Bergh\"ofer T.W., Boer B. 1998, A\&A 330, 823
\bibitem[1990]{stockton}
Stockton A. 1990, in "Dynamics and Interaction s of Galaxies", ed. Wielen R., Springer Verlag, Berlin, p.440
\bibitem[1994]{suchkov}
Suchkov A.A., Balsara D.S., Heckman T.M., Leitherer C. 1994, ApJ 430, 511 (S94)
\bibitem[1988]{telesco}
Telesco C.M. 1988, ARAA 26, 343
\bibitem[1992]{thuan}
Thuan T.X., Sauvage M. 1992 A\&AS 92, 749
\bibitem[1986]{tomisaka1}
Tomisaka K., Ikeuchi S. 1986, PASJ 38, 697
\bibitem[1988]{tomisaka2}
Tomisaka K., Ikeuchi S. 1988, ApJ 330, 695
\bibitem[1989]{turatto}
Turatto M., Capellaro E., Petrosian A.R. 1989, A\&A 217, 79
\bibitem[1993]{turner}
Turner T.J., George I.M., Mushotzky R.F. 1993, ApJ 412, 72
\bibitem[1983]{weedman}
Weedman D.W. 1983, ApJ 266, 479
\bibitem[1985]{woosley}
Woosley S.E., Weaver T.A. 1985, in "Radiation Hydrodynamics in Stars and Compact Objects", Proc. of the IAU Colloquium No. 89, eds. Mihalas D., Winkler K.H., p.91
\bibitem[1997]{zimmermann}
Zimmermann H.U., Becker W., Belloni T., et al. 1997, EXSAS User's Guide, Edition 5, MPE Report, Garching 
\end{thebibliography}
\end{document}